# Einstein's conversion from his static to an expanding universe

Harry Nussbaumer
Institute of Astronomy
ETH Zurich
CH-8093 Zurich
Switzerland

**Summary**
In 1917 Einstein initiated modern cosmology by postulating, based on general relativity, a homogenous, static, spatially curved universe. To counteract gravitational contraction he introduced the cosmological constant. In 1922 Alexander Friedman showed that Albert Einstein's fundamental equations also allow dynamical worlds, and in 1927 Georges Lemaître, backed by observational evidence, concluded that our universe was expanding. Einstein impetuously rejected Friedman's as well as Lemaître's findings. However, in 1931 he retracted his former static model in favour of a dynamic solution. This investigation follows Einstein on his hesitating path from a static to the expanding universe. Contrary to an often advocated belief the primary motive for his switch was not observational evidence, but the realisation that his static model was unstable.

**1. Introduction**

It has become a popular belief that Albert Einstein abandoned his static universe when, on a visit to Pasadena in January and February 1931, Edwin Hubble showed him the redshifted nebular spectra and convinced him that the universe was expanding, and the cosmological constant was superfluous. "Two months with Hubble were enough to pry him loose from his attachment to the cosmological constant" is just one example of such statements [Topper 2013]. There are variations of the theme, but the essence remains the same: Hubble personally convinced Einstein that the universe was in a state of expansion.

The present investigation shows that this stereotype misses the truth by a large margin. We shall first recall the very early days of modern cosmology and then follow the key events surrounding the discovery of the expanding universe, in which Einstein took no part. He most probably became aware of the progress in cosmology when in June 1930 he visited Arthur S. Eddington. We then follow him on his journey to Pasadena in January and February 1931, where close contacts with Richard C. Tolman must have helped him to come to grips with the new developments in cosmology. After his return to Berlin he wrote and submitted in April 1931 a report to the Prussian Academy of Sciences, where he adopted a model of the expanding universe proposed by Alexander Friedman in 1922. In January 1932 Einstein was back in Pasadena, where he met Willem de Sitter, with whom he teamed up to write what would be known as the Einstein-de Sitter universe, which up to the middle of the 1990s was the generally accepted cosmological model.

## 2. The beginning of modern cosmology: Einstein's static universe

Modern cosmology began in 1917 with Einstein's "Kosmologische Betrachtungen zur allgemeinen Relativitätstheorie" (Cosmological considerations on the general theory of relativity) [Einstein 1917]. He applied general relativity to the entire universe. To him it must have been a matter of common sense that we lived in an immutable cosmos; thus, theory had to describe a static universe. However, his original field equations did not give such results. If matter was homogenously distributed and gravitation was the only active force, his universe would collapse. He therefore introduced the famous cosmological term, λ, today usually designed by Λ, so that his fundamental equations took the form

$$(G_{ij} - \lambda g_{ij}) = -\kappa \left( T_{ij} - \frac{1}{2} g_{ij} T \right). \quad (1)$$

κ is Einstein's constant of gravity. The left hand side describes the geometrical structure of the universe, the right hand side represents the energy-momentum tensor due to the action of matter; the λ-term acts as a repulsive force. This additional term gave Einstein a static, spherical, spatially closed universe. He emphasised that the known laws of gravitation did not justify the introduction of λ, its inclusion was motivated by the quest for a static solution of the differential equations. There is a simple relationship between λ, κ, the mean material density ρ, and the radius of curvature $R$:

$$\lambda = \frac{\kappa \rho}{2} = \frac{1}{R^2}. \quad (2)$$

The total mass of Einstein's spatially closed universe can be calculated as $M = \rho 2\pi^2 R^3$. Already in 1916, in a footnote of his review article in the *Annalen der Physik*, reprinted in CPAE [Einstein 1916, footnote on page 319], Einstein had considered the introduction of λ; but then he dropped it without giving any reason.

A few months later Willem de Sitter published an alternative model, which also was supposed to represent a static universe. However, de Sitter's world was devoid of matter, thus, the right-hand side of equation (1), was zero [de Sitter 1917]. Einstein was not happy with de Sitter's solution, where, in the absence of matter λ took sole responsibility for the physical properties of the universe. Further details on the cosmological discussion of the early 1920s can be found elsewhere (e.g. [Nussbaumer and Bieri 2009, chapter 6], [Jung 2005]).Let us emphasise that Einstein never really came to terms with his brainchild λ. We find an early example in 1923. In an exchange of letters and postcards with Hermann Weyl, Einstein pointed out that de Sitter's world was not truly static, because two test particles would rush apart, and he added "if there is no quasi-static world, then away with the cosmological term" [Einstein 1923c; Nussbaumer and Bieri 2009, chapter 6 shows the postcard].

## 3. Einstein's reaction to Friedman's dynamic universe

Einstein's first confrontation with a truly dynamic universe was Friedman's 1922 article "Über die Krümmung des Raumes" (About the curvature of space) [Friedman 1922]. Based on Einstein's fundamental equations of general relativity he retained Einstein's hypothesis of a homogenous, isotropic 4-dimensional, positively curved universe, and he also kept the cosmological constant, λ. As a fundamental difference against Einstein and de Sitter, Friedman allowed the radius of curvature, $R$, to be variable with time.

The difference between the models of Einstein, de Sitter, and Friedman can best be shown with the line element which defines the relation between the spatial fraction of the 4-dimensional spacetime with time as fourth dimension. Friedman writes it as

$$ds^2 = R^2 \left( dx_1^2 + \sin^2 x_1 dx_2^2 + \sin^2 x_1 \sin^2 x_2 dx_3^2 \right) + M^2 dx_4^2 \quad . \quad (3)$$

He recovered Einstein's and de Sitter's worlds when he replaced $R^2$ by $-R^2/c^2$, and set $M=1$ in the model of Einstein and M=cos($x_1$) in that of de Sitter. In Einstein's and de Sitter's models $R$ is a constant. Friedman allows $R$ to vary with time: $R=R(x_4)$. If we choose $x_2=0$ and $x_3=0$ then for a beam of light, $ds=0$, we have $dx_4= (R/c)·dx_1$, whereas for de Sitter's model we find $dx_4= \cos(x_1)·(R/c)·dx_1$. From Einstein's fundamental equations Friedman then derived equations for the determination of $R$ and the matter density ρ, both of which are allowed to vary with time:

$$\frac{R'^2}{R^2} + \frac{2RR''}{R^2} + \frac{c^2}{R^2} - \lambda = 0, \quad (4)$$

and

$$\frac{3R'^2}{R^2} + \frac{3c^2}{R^2} - \lambda = \kappa c^2 \rho. \quad (5)$$

These equations will serve Einstein, when in 1931 he finally accepts the dynamic universe [Einstein 1931e], as well as Einstein and de Sitter, when in January 1932 they proposed the Einstein-de Sitter model [Einstein and de Sitter 1932].

In Einstein's universe of 1917 time, $x_4$, is running everywhere at the same speed, whereas in de Sitter's universe of 1917 time runs differently for different values of $x_1$. This resulted in a redshifted spectrum for remote sources; this was a strong point of de Sitter's model, because it seemed to explain Slipher's redshifts observed in distant extragalactic nebulae [e.g. Slipher 1917]. In 1925 Lemaître showed that de Sitter's model violated the basic cosmological principle of spatial homogeneity [Lemaître 1925].

Only Einstein responded to Friedman's publication. His responses are well known [Nussbaumer and Bieri 2009, chapter 7]. In his first reaction he discarded Friedman's solution as not compatible with his field equations [Einstein 1922]. Einstein then realised his error. In the draft of his second answer he acknowledged that formally Friedman was right, but that his dynamical solutions were hardly of physical significance [Einstein 1923a]. In the published version Einstein deleted that comment [Einstein 1923b], but it obviously corresponded to his true conviction, as his later reaction to Lemaître showed. Friedman's article was purely theoretical, without any references to particular observations.

## 4. Lemaître discovers the expanding universe and talks with Einstein

Einstein's second confrontation with a dynamic universe came in the autumn of 1927. Lemaître, who did not know the work of Friedman, had taken a fresh look at Einstein's fundamental equations [Lemaître 1927]. And, just as Friedman before him, he kept the cosmological constant and allowed the radius of curvature, $R$, as well as the density, ρ, to vary with time. Unlike Friedman, Lemaître spotted the weak point in de Sitter's model of 1917: it violated the principle of spatial homogeneity.

From his theoretical, dynamical model he concluded that spectra of distant, extragalactic nebulae would allow one to distinguish between static, contracting and expanding worlds. For an expanding universe redshifted spectra had to be expected with a tight relationship between nebular distances and redshifts of the form

$$v = H \cdot d, \quad (6)$$

if redshifts were treated as Doppler-shifts corresponding to a velocity $v$; $d$ is the nebular distance and $H$ a factor of proportionality, which Lemaître determined from observations and which later was named the Hubble constant. Lemaître was well aware of the observational status of cosmology. He knew Hubble's distances to the extra-galactic nebulae [Hubble 1926], and he knew Slipher's measured redshifts, which he had found in Strömberg's publication [Strömberg 1925]. Although nebular distances were still loaded with great uncertainties, he concluded that the available observations were compatible with his theoretically derived relationship of equation (6). This allowed him to postulate an expanding universe.

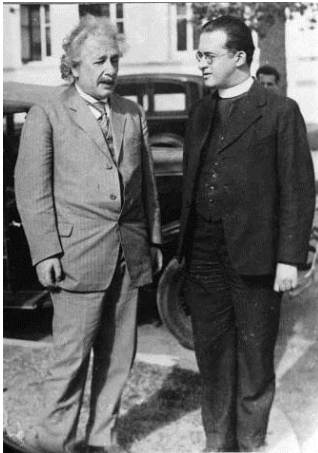

**Einstein and Lemaître.** Photographed around 1933. (Archives Lemaître, Université Catholique, Louvain.)

Lemaître published his discovery in a little known Belgian scientific journal [Lemaître 1927]. The discovery went unnoticed. But, on the occasion of the 1927 Solvay congress Lemaître handed a reprint to Einstein. After reading it, the already legendary Einstein and young Lemaître discussed the article: "Après quelques remarques techniques favorables, il conclut en disant que du point de vue physique cela lui paraissait tout à fait abominable." (After a few favourable technical remarks he concluded by saying that from a physical point of view this looked to him abominable.) It was probably on this occasion that Lemaître learnt about Friedman's earlier research.

During the taxi ride to Auguste Piccard's laboratory Lemaître explained to Einstein the observational status of cosmology. He came away with the impression that Einstein was not at all informed about the astronomical facts [Lemaître 1958].

Thus, Einstein, even when confronted with observational evidence that the universe was a dynamic and not a static entity, still refused to consider such a cosmology as an alternative to his static world.

**5. Einstein visits Eddington**

In 1929 Hubble published his observational finding that there was probably a linear relationship between nebular distances and redshifts [Hubble 1929]. Hubble was not aware that he had verified Lemaître's prediction of 1927. The excitement among theoreticians was high, as shown by the report in the *Observatory* about a Royal Astronomical Society meeting on Friday, 10 January 1930, where de Sitter presented his own investigation and confirmed Hubble's announcement. However, Eddington and de Sitter were at a loss how to theoretically interpret this observational fact. The minutes of the meeting were published in the February issue of the *Observatory* [de Sitter 1930a]. Lemaître read them and immediately sent two reprints of his 1927 paper to Eddington, reminding him that he should have received a reprint in 1927, and begging him to send the second copy to de Sitter. Eddington read and saw immediately that Lemaître had found the solution to the enigmatic redshifts and the accompanying cosmological problem. Also de Sitter was immediately convinced of Lemaître's solution.

In their ensuing publications both acknowledged Lemaître's breakthrough and adopted his model of an expanding universe. In the May issue of the *Monthly Notices* Eddington published an article entitled "On the instability of Einstein's spherical world" [Eddington 1930]. It begins with a generous recognition of Lemaître's discovery: *"Working in conjunction with Mr. G.C. McVittie, I began some months ago to examine whether Einstein's spherical universe is stable. Before our investigation was complete we learnt of a paper by Abbé G. Lemaître which gives a remarkably complete solution of the various questions connected with the Einstein and de Sitter cosmogonies. Although not expressly stated, it is at once apparent from his formulae that the Einstein world is unstable – an important fact which, I think, has not hitherto been appreciated in cosmogonical discussions."* He then says that in connection with the behaviour of spiral nebulae he had hoped to contribute in this respect some definitely new results, but this "*has been forestalled by Lemaître's brilliant solution*". In the same publication Eddington showed that Einstein's universe was unstable [Eddington 1930]. Lemaître's findings and Hubble's observational confirmation of the predicted velocity-distance relationship signalled the demise of the static universe.

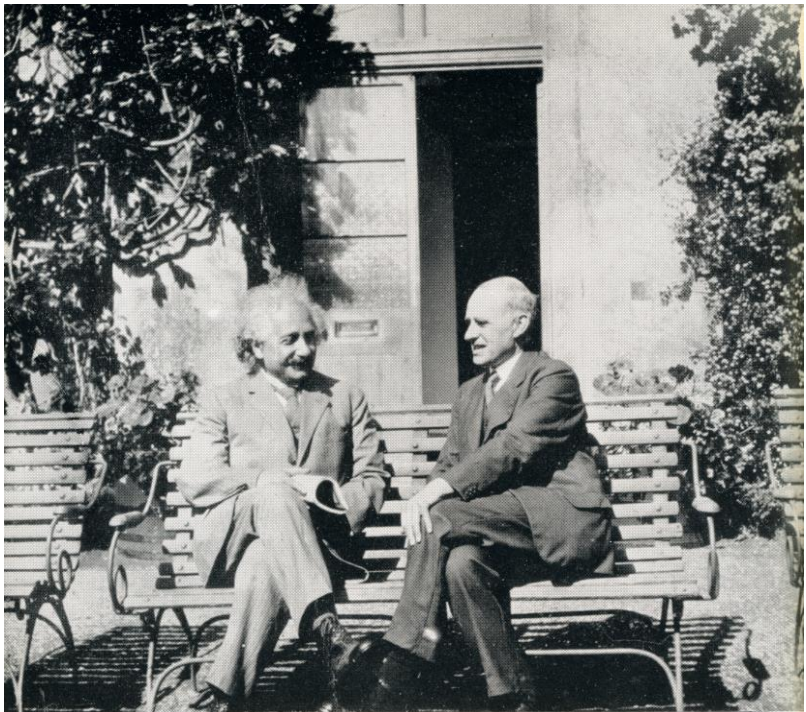

**Einstein (1879-1955) and Eddington (1882-1944) in Cambridge**. In June 1930 Einstein stayed with Eddington and his sister, who lived in the Director's apartments on the east side of the Observatory. In the background the French doors of Eddington's ground floor study (Information from Simon Mitton). Photograph by Miss W. Eddington [Douglas 1956].

In June 1930 Einstein visited Eddington. Vibert Douglas' biography of Eddington briefly records the visit and gives a photograph of Einstein and Eddington, with whom Einstein stayed [Douglas 1956]. At that time Eddington was one of the most prominent astrophysicists. He was a highly competent theoretician, and with his participation in the solar eclipse observations, where a fundamental issue of general relativity was verified, he had much contributed to Einstein's fame [Dyson et al. 1920]. In June 1930 neither Eddington nor Einstein kept a diary. But even without written proof it is unthinkable that in their conversation they would not have discussed the cosmological issue. Einstein most likely had no clue about the observational situation, other than that described to him by Lemaître in 1927, whereas Eddington had in 1923 included Slipher's redshifts in his *Mathematical Theory of Relativity* [Eddington 1923/1924], and he knew about Hubble's and de Sitter's investigations into the redshift-distance relation. Eddington formed with de Sitter the most competent acknowledged pair in cosmology. We may confidently assume that in Eddington's home Einstein was updated on observational and theoretical cosmology. Thus, when at the beginning of December 1930 Einstein sailed for Pasadena, he was at least superficially aware of what was going on in that field. However, when he arrived in Pasadena his mind was in a bewildered state. Observations were hardly the problem, de Sitter vouched for their validity [de Sitter 1930b], but giving up the notion of a static universe was another matter.

**6. Einstein's first journey to Pasadena**

In 1914 Einstein left his ETH-professorship in Zurich and settled in Berlin, where he was a member of the „Preussische Akademie der Wissenschaften zu Berlin" and since 1917 he also acted as director of the *Kaiser-Wilhelm-Institut für Physik*. Millikan and his colleagues had for years tried to have Einstein at their institute for some limited time. Whereas in 1928 his weak heart severely limited his activities, in 1930 he finally accepted to travel. In a press release he stated the reason and purpose of his journey: *Ich fahre nach Pasadena auf Einladung der dortigen Universität hin. Der Hauptzweck der Reise ist die Teilnahme am dortigen wissenschaftlichen Leben, der Verkehr und Diskussionen mit den Fachkollegen. Vielleicht werde ich auch einige Vorträge zu halten haben* [Einstein 1930]. (I am going to Pasadena at the invitation of the University there. The main reason for the trip is participation in their scientific life and contact and discussion with colleagues in my disciplines. Perhaps I'll also have to give some lectures.)

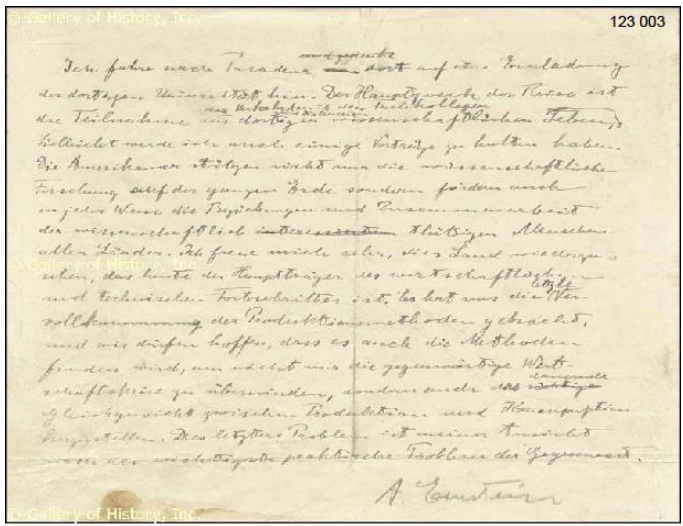

**Press release December 1930.** Einstein gives his reason for travelling to Pasadena [Einstein 1930].

His journey may also have been triggered by his increasing disillusionment about Germany's political future. We might suspect that already in 1930 the journey to Pasadena was a welcome opportunity to explore future possibilities for a job. On December 3 the sea journey began towards Southampton and continued on the following day to America.

*Einstein at sea*
During the journey to Pasadena Einstein kept a diary into which he irregularly entered his activities [Einstein 1930/1931]. It begins with his departure by train from Berlin, Bahnhof Zoo, in the evening of November 30, 1930, accompanied by his wife Else, seen off by relatives, photographers, and reporters. Arrival in Köln the following morning at 8 a.m., change of trains and arrival in Lüttich at 11 a.m. There they were met by Einstein's secretary Helen Dukas and by the mathematician Walther Mayer (1887-1948), who was Einstein's collaborator. We now extract those entries that mention his scientific preoccupations. This allows us to situate his thinking about cosmology within his wider scientific activities.

On December 2, the party embarked on the Schelde to reach their ship for crossing the ocean *Während Scheldefahrt wissenschaftliche Unterhaltung mit Mayer über den Stand unseres Problems und Prinzipielles über Quantentheorie.* (During the Schelde trip scientific discussion with Mayer about the status of our problem and on principal points of quantum theory.)

On December 4 he reports that in spite of lovely weather, fish jumping above the water, and reporters pestering him orally and via telegraph: *Ich aber rechne am Bewegungsgesetz.* (But I work on the law of motion.)

*8. XII. […] Mayer findet viele Gleichungen mit einer Identität. Man sieht, dass Hamilton'sches Prinzip grosse auswählende Kraft hat. Studiere fleissig Quanten-Mechanik. Empfinde sie als unnatürlich, bewundere aber den Scharfsinn. Gute Theorie muss aus Feld-Idee herauswachsen. Mayers Konsequenz und Ordnung im Arbeiten ist aussergewöhnlich.* (Mayer finds many equations with an identity. One sees that the Hamiltonian principle has great selective power. Busy studying quantum mechanics. Perceive it as unnatural, but admire its ingenuity. Good theory must grow from the field-idea. Mayer's consistency and order in work are extraordinary.)

*9. XII. […] Feldgleichungen durch Variation der antiparallelen Krümmung gebildet und wieder verworfen.* (Field equations built through variation of the antiparallel curvature but then discarded.)

Einstein arrives in New York on December 11 and comments on the huge publicity surrounding his arrival – photographers that lunge out at him like hungry wolves. On December 14 he is relieved to be back on sea.

*15. […] Ich arbeite viel, mit und ohne Mayer, schreibe eine Darstellung der allg. Rel. Feldtheorie, die an Universität verkauft wird.* (Work a lot with and without Mayer, write an exposition on relativistic field theory which will be sold to university.) He gives no further information about that project of presenting the field theory.

The sea journey continued via Cuba, with a stop in Havana, then through the Panama channel with a visit to Panama's president, who had been studying in Zurich.

*26. XII. […] Gestern Form der Feldgleichungen ( $A^s_{\mu\nu\nu} = 0$ ) gefunden mit Koeffizientenbestimmung aus der Bedingung der Singularitätsfreiheit. […]* (Yesterday found form of field equations ( $A^s_{\mu\nu\nu} = 0$ ) with coefficients found from the condition of freedom from singularities.)

*29. XII […] Mayerchen und ich haben fleissig Dirac studiert. Die Quanten-Mechanik ist ein wirklich geistreiches System.* (Mayerchen [diminutive of Mayer] and I were busy studying Dirac. Quantum theory is a truly ingenious system). – Einstein reports about work on general relativity, on quantum mechanics, and that he studies Dirac, but he never mentions cosmology.

**7. Einstein in Pasadena**

The famous professor Einstein was, of course, for the journalists a much sought after prey. But his presence was also precious publicity for the Mount Wilson Observatory and the California Institute of Technology and they made good use of it.

*An interview with The New York Times*
The ship arrived in San Diego in the early morning of December 31. The city welcomed Einstein with a big reception. In the afternoon Fleming took him to his house in Pasadena, where he could relax. On January 2, 1931 the diary reads: *Gang ins Institut. Karman, Epstein, und hiesige Kollegen. Nachmittags zuhause. Rundgang im Garten.* (Went to the Institute Karman, Epstein and collegues from here. Afternoon at home. Stroll in the garden.)

Which are Einstein's scientific priorities in Pasadena, and what is his cosmological picture at the beginning of his stay? On January 3, 1931 the New York Times published an interview with Einstein that took place January 2 [Einstein 1931a]:
*Dr. Albert Einstein disclosed today why he came to California.*
    *He expects help from the scientists at Mount Wilson Observatory and the California Institute of Technology to solve the major problem of his mind, whether gravitation, light, electricity and electromagnetism are not different forms of the same thing.*
*"This question is the main problem of the present relativity theory," he explained. "Interesting possibilities have been found to solve the question, but it is still uncertain if the ways being tried will be successful.*
    *"New observations by Hubble and Humason (astronomers at Mount Wilson) concerning the red shift of light in distant nebulae make the presumptions near that the general structure of the universe is not static.*
    *"Theoretical investigations made by La Maitre and Tolman (mathematicians of California Institute of Technology) show a view that fits well into the general theory of relativity."*
The interview goes on, but cosmology is no longer the subject and we can stop here.

That interview made it to the front page of the New York Times on January 3, 1931. The unification of general relativity, electromagnetism and quantum mechanics into one single theory is foremost in his mind. The reference to Mount Wilson and gravitation may indicate his interest in the solar observations of Charles Edward St. John, who conducted observational tests on the validity of Einstein's theory of a gravitational redshift in the solar spectrum. But he also mentions cosmology: Observations by Hubble and Humason, and two theoreticians whose explanations, based on general relativity, fit the observations. The journalist must have misunderstood: One of them was indeed Richard C. Tolman of the California Institue of Technology, but the one called "La Maitre" was really Georges Lemaître from the Université catholique de Louvain (Belgium), who in 1927 had discovered the expansion of the universe. Eddington, in his 1930 article on the instability of Einstein's universe, mentions them both; he highly praises Lemaître, but doubts Tolmans's solution.

Although we have no detailed knowledge on the source of Einstein's information, we know that in 1927 Lemaître had confronted him with the nebular redshifts. And in June 1930, thus just six months prior to his arrival in Pasadena, Eddington had certainly informed him about Hubble's observationally determined linear relationship between distance and redshift of the nebulae, a relationship predicted by Lemaître.

*Tolman, the theoretician at Pasadena*
In his book *Die Welt als Raum und Zeit*, published in Russian in 1923, Friedman was the first to speculate within the frame of general relativity about the origin of the present universe [Friedman 1923]. However, Friedman had no observations to judge the plausibility of his scenarios. By the end of the 1920s the situation was different. Tolman was an eager contributor to the cosmological debate. He was certainly aware that the British cosmologists Eddington, McCrea, and McVittie were investigating, whether the present expansion could have begun due to inhomogeneities out of an Einstein-equilibrium. The result was not encouraging: *a universe containing a single condensation would start contracting* [McCrea and McVittie 1930]. It showed that there was no easy explanation for the existence of an expanding universe. Lemaître's suggestion of a "Big Bang" would only be published in May 1931: *We could conceive the beginning of the universe in the form of a unique atom, the atomic weight of which is the total mass of the universe. This highly unstable atom would divide in smaller and smaller atoms by a kind of super-radioactive process* [Lemaître 1931].

But what was Tolman's position and attitude? In an article, submitted in August 1930 and published in September 1930, he tells us: *In several previous articles, I have shown the possibility of deriving a non-static cosmological line element to agree with the relatively uniform distribution of matter observationally found in the universe, and of using this line element to account both for the annihilation of matter taking place throughout the universe, and for the red-shift in the light from the extra-galactic nebulae. After the publication of the first of these articles, I learned from a conversation with Professor H. P. Robertson that he had previously published a different derivation of the same general form of line element; more recently Professor de Sitter and Sir Arthur Eddington have both been kind enough to call my attention to the previous use of a non-static line element by Lemaitre, and still more recently I have discovered the early treatment of non-static line elements by Friedman.*

For Tolman annihilation of matter had become a central theme in cosmology [see Nussbaumer and Bieri 2009, chapter 13]. Anyway, Tolman was well informed on the cosmological front, and for him and Einstein there would be no lack in topics of conversation. (Tolman understood German very well).

*What Einstein's diary tells us*
The New York Times interview is silent on Einstein's personal attitude to the proposed theoretical models. However, it is worth noting that we hear no anti-dynamic statement. His stubbornly static attitude had most probably been shaken during his visit to Eddington. We now dive again into Einstein's diary to follow his scientific preoccupations. – The diary shows that Einstein was extremely busy. He could not avoid masses of social engagements, and his activities often touched matters of public interest. He openly and strongly advocated conscientious objection, was consulted on educational issues, and was often approached by the Jewish community for moral support.

*3. I. 31. Arbeiten im Institut. Zweifel an Richtigkeit von Tolmans Arbeit über kosmologisches Problem. Tolman hat aber Recht behalten.* (Work at the Institute. Doubts about the correctness of Tolman's work on the cosmological problem. But Tolman is right.) In 1929 and 1930 Tolman had published five papers with different suggestions on the cosmological problem, we do not know which of these publications Einstein was referring to, nor on which point Tolman was right. The immediate preoccupation with Tolman's ideas underlines that for Einstein the cosmological problem was a predominantly theoretical issue.

*7. Gestern Abend bei Millikan, der hier die Rolle des lieben Gottes spielt. Heute grosse Kinoaufnahme mit Millikan, Michelson, Adams. Heute war astronomisches Kolloquium Sonnenrotation v. St. John. Sehr sympathischer Ton dort. Habe wahrscheinliche Ursache der Veränderlichkeit der Sonnenrotation gefunden in Zirkulationsbewegung, die von Abplattung und durch diese bewirkte relative Abkühlung am Aequator gefunden. Heute trug ich über Gedankenexperiment [… ?] vor im theoret. physik. Kolloquium. Gestern war physikalisches Kolloquium über Einfluss von Magnetfeld bei der Krystallisation auf die Eigenschaften der Wismuth-Krystalle.* (Last night with Millikan who plays here the part of our Good Lord. Today great motion-picture taking with Millikan, Michelson, Adams. Today was an astronomical colloquium on solar rotation by St. John. Very congenial atmosphere. I probably found the cause of the variability of the solar rotation in the circulation due to cooling at the equator because of the oblateness. Today I spoke about thought experiments [… ?] at the theoretical physics colloquium. Yesterday physics colloquium on the influence of the magnetic field at the crystallisation on the properties of bismuth-cristals.)

*9. […] Mittags erlaubt mir der Arzt Besuch von Mount Wilson. Gestern hübsche Idee über Ursache der Zirkulation an der Sonnenoberfläche. Abends Besuch bei Millikans. Verfassung einer Tischrede für den 15. über hiesige Wissenschaftler.* (In the afternoon the doctor allowed me to visit Mount Wilson. Yesterday pretty idea on the cause of the circulation on the solar surface. In the evening visit at the Millikan's. Composition of an after-dinner speech about local scientists.)

*22. I. Lange Pause aber viel dazwischen. […] Das tollste war ein Abend für die finanziellen Gönner des Instituts. 350 Händedrucke mit salbungsvollen Radio-Reden, auch von mir. Abend des Un. Klubs mit Relativitäts-Rede St. Johns. Wissenschaftlich interessante Dinge, Astronomisch die Platten der Doppler-Verschiebungen von Nebeln, Gespräch mit St John über Rotverschiebungen in der Sonne Magnetfelder in Sonnenflecken in aufeinanderfolgenden Sonnen-Perioden verschiedenes Vorzeichen Bleibt Mysterium. – Vortrag des vortrefflichen Tolman über Relativitäts Thermodynamik. Leichte Grippe mit Fieber überstanden. Mayer findet Verbesserungen zu Dirac's Quantenmechanik. […] Einen interessanten Vortrag Karmans über Turbulenzen sowie einen Bericht Zwickys über Sekundär-Struktur in Krystallen darf ich auch nicht vergessen. Ich finde es sehr interessant mit den Kollegen, die anregend und wirklich freundschaftlich sind. Nur komme ich zu wenig zum Selber-Arbeiten.* (Long break but a lot in between. […] The most fantastic was an evening for the financial sponsors of the Institut. 350 handshakes with unctuous radio-speeches, also from me. Evening of the Un. club with relativity speech by St. John. Scientifically interesting things, astronomical the plates with the Doppler-shifts of nebulae, talk with St. John about redshifts in the sun, magnetic fields in sunspots in successive solar-periods different sign remains mystery. – Talk of the excellent Tolman about relativistic thermodynamics. Light influenza with temperature well overcome. Mayer finds improvements to Dirac's quantum-mechanics. […] I must not forget an interesting talk by Karman about turbulence as well as a report by Zwicky about secondary structure in crystals. I find it very interesting with the collegues, who are truly stimulating and friendly. Only, I find too little time to work for myself.)

### The Los Angeles Times on Hubble and Einstein

The front page of the Los Angeles Times of January 24, 1931 highlights a lecture given on January 23 by *Dr. Edwin P. Hubble, Mt. Wilson astronomer and Einstein collaborator, who addressed a distinguished group of scientists in the Mt. Wilson Observatory laboratory late today. Dr. Hubble and his associate, Milton G. Humason, conferred before the lecture with Dr. Albert Einstein, who plans to leave this city for Palm Springs tomorrow afternoon. […] Tonight the author of the new unified-field equation attended a motion-picture preview in Los Angeles, according to unofficial reports.*

*The Mt. Wilson astronomer, who has explored farther toward the outermost limits of creation than any other living man, revealed that from a study of 1400 photographic plates taken with the telescope at Mt. Wilson and at the Lick Observatory he has counted 30,000,000 nebulae, or island universes, each of which contains billions of suns and planets. The major part of his discussion was an explanation of the method by which he has calculated that these millions of universes are approximately equally distributed in space. […]*

*"We must suppose" Dr. Hubble asserted, "that creation itself extends far beyond the limits of the most powerful telescope that has been built and that these inaccessible regions are similar to those already observed".*

*Dr. W.W. Campell, former president of the University of California and noted Lick observatory astronomer, and Dr. Robert, A. Millikan, executive head of the California Instutute of Technology, were present and took part in the discussion following Dr. Hubble's talk.*

*Dr. Einstein, who has indicated that creation might be measured by its density or the amount of water it contains, has expressed great interest in the work of Dr. Hubble, who recently announced an estimate of the extent of "everything" based upon his study of distant nebulae* [Los Angeles Times January 24, 1931].

On January 25, Einstein's diary does indeed mention his trip to Palm Spring: *Heute hatten wir eine wundervolle Fahrt in kalifornische Wüste nach Palm Spring* (Today we had a wonderful trip into [the] Californian desert to Palm Spring). But the diary does not mention Hubbel's lecture. There are two more Pasadena-entries: On January 26 and 27 he writes about sightseeing trips, but then jumps to April 8, when he was back in Germany.

### *No mentioning of Hubble in Einstein's diary*
We are astonished: Einstein's diary of his first journey to Pasadena does not mention Hubble at all. Yet, the newspaper just told us that they met, and on other occasions we see them together on photographs, and the diary mentions the nebular plate collection, it would be surprising if Hubble had not been present on that occasion. He mentions lectures and meetings with other scientists, yet Hubbel's name is nowhere. Obviously, meeting Hubble did not particularly impress Einstein and did definitely not trigger any new insights. Having a look at the collection of spectral plates was in his case hardly more then a scientific tourist attraction, in themselves they are not spectacular and on their own they don't prove anything. Einstein had accepted the observational fact of the nebular redshifts well before his arrival on Mount Wilson, after all, de Sitter had given his blessing. What he knew about observations he had most probably learnt from Eddington, supplemented by Tolman, who knew German, whereas we have no reliable evidence that Hubble knew German sufficiently well to join in a conversation. Equally important might have been Hubble's reluctance to get involved in theory. In a letter to de Sitter about redshifts Hubble stated in 1931: *The interpretation, we feel, should be left to you and the very few others who are competent to discuss the matter with authority* [see Nussbaumer and Bieri 2009, chapter 12]. It is also significant that in his lecture, as reported by the Los Angeles Times, Hubble explained the methods of his observations and the distribution of the nebulae in space, however, he did not touch the subject of the redshifts and their possible cosmological implications.

This surprising silence on Hubble does not imply that Einstein attached little value to observations or Hubble's contribution. Einstein was not the stereotype theoretician with his mind in the clouds, he had started his professional career at the Swiss Patentamt, and he knew that Slipher's redshifts had since 1917 been considered a strong support for de Sitter's model. But now, that de Sitter had abandoned his model, and his own static world had been shown to be unstable, whereas Lemaître's predicted velocity-distance relationship had been verified by Hubble, he was forced to reconsider his former commitment to the immutability of the heavens. From Hubble no further enlightenment was to be expected. At Pasadena his man for cosmology was Tolman and not Hubble. The Einstein Archive contains a sheet of paper with a dedication from Einstein to Tolman, testifying to a close and friendly relationship between the two. – Einstein's interest in observations seems to have been more directed to the solar observations of Mount Wilson, rather than to Hubble's nebulae. They pertained to the gravitational redshift in the solar spectrum, predicted by general relativity.

### *Will the unified field theory solve the cosmological problem?*
On February 5, 1931 the Los Angeles Times carried a report by a staff correspondent about an astronomy seminar given by Einstein on February 4 [Los Angeles Times February 5, 1931]: *Futuristic art enthusiasts may derive satisfaction from Dr. Albert Einstein's revelation today that the curves of space, as indicated by his new unified field theory equation, are less beautiful than the curves suggested by the general theory of relativity. In fact, some scientists who heard Dr. Einstein speak in German at an astronomy seminar at the Mt. Wilson Observatory laboratory, interpreted his closing words to mean that as a result of his most recent work he has discovered that creation no longer may be conceived as perfectly inclosed or finite, as formerly suspected. [...]*

  Einstein was asked to explain the relation of the unified field theory with cosmology: *His answers, limited to a few sentences, indicated, according to translators, that the first major result of the new theory will be a modification of his former idea that space is inclosed.*

  *"Regardless of what equation is used," he is reported to have said, "space cannot be closed in just the Euclidian sense, but we must add something to this."*

  *Dr. Walter S. Adams, Mt. Wilson Observatory director, and Dr. Charles E. St. John and Dr. Edwin P. Hubble, Einstein scientific collaborators, who were present, voiced the opinion that the German theoretical physicist did not have time fully to explain his closing statement and cautioned the press against leaping to premature conclusions.* It was agreed that a further meeting should be arranged, so that Einstein could further elaborate on *the possible revolutionary effect of his newest equation, recently announced here, upon scientific conceptions of creation.*

  The report finishes with a statement of the opinions regarding cosmology: *Prior to the all-inclusive theory and its now apparent revolutionary effect upon conceptions of the universe, there have been three opposed general cosmic interpretations.*

  *Dr. Einstein himself stressed matter, supposing a finite density of matter. Dr. W. de Sitter opposed this view with the supposition that creation tended to be empty. More recently, Dr. Richard Tolman of the California Institute of Technology brought these two previous views of the cosmos nearer together with his postulate of a nonstatic universe – a creation in which matter is continuously being radiated into energy.*

  *When Dr. Einstein amplifies his bare allusion to the effect of his newest theory upon these opposing views, local scientists anticipate that a fourth and far completer picture of creation will be hung in the art galleries of modern scientific speculation.*

This report suggests that for a short while Einstein might have hoped to solve the cosmological problem with the unified field theory he was working on. Tolman's cosmological involvement we shall further discuss in the following section.

### *The New York Times reports a meeting of February 11, 1931*
Einstein's stay at Pasadena was drawing to a close. The New York Times reports a meeting on February 11, 1931; it was focused exclusively on the cosmological issue [Einstein 1931b].
*Dr. Albert Einstein told astronomers and physicists here today that the secret of the universe was wrapped up in the red shift of distant nebulae.*

*For more than an hour he discussed possibilities of the shape of the universe to take the place of his discarded Einsteinian universe. The meeting, held at the Mount Wilson Laboratory, was attended by the Carnegie Institution of Washington's astronomers who discovered this red shift, Dr. Edwin C. Hubble and Dr. Walter S. Adams.*

This was probably the follow up meeting to the discussion on February 4, reported in the Los Angeles Times. The press coverage was doubtless welcome to the directorate of the Mount Wilson observatory, whose director W.S. Adams saw a further opportunity to attract public attention to his institution. What information they gave to the press before or after the meeting we do not know, but we can read that the journalists were left with the impression that Hubble and Adams had discovered the nebular redshifts. This is a far cry from the facts. Nebular redshifts were discovered by Slipher as early as 1912 [Slipher 1913]. Hubble, in his 1929 publication, based his velocities mostly on Slipher's redshifts [Nussbaumer and Bieri 2009, chapters 5 and 10; Peacock 2013; O'Raifeartaigh 2013].

The report continues:
*"The redshift of distant nebulae has smashed my old construction like a hammer blow," said Einstein, swinging down his hand to illustrate. "The red shift is still a mystery."*
*"The only possibility is to start with a static universe lasting a while and then becoming unstable and expansion starting, but no man would believe this."*
Einstein delivers an impressive punch line. However, his statement is basically still the same as in the interview of January 3: observations suggest that the universe is not static, however these redshifts still remain a mystery. But he then continues with an interesting declaration. He takes up Eddington's favourite explanation of the world's past history: the universe has existed since eternity in an Einsteinian pseudo-static state, but at some moment in the past it began to slide into expansion [Eddington 1931]. During 1930 and 1931 this topic was debated by Eddington and his group and also by Lemaître; for details see Nussbaumer and Bieri [Nussbaumer and Bieri 2009, chapter 17]. One result of the debate was Lemaître's suggestion in 1931 of the "Big Bang" [Lemaître 1931]. Einstein must have heard of the discussions during his 1930 visit to Eddington, and he certainly discussed the matter with the "excellent Tolman", who was aware of the debate. Einstein qualifies Eddington's favourite model as: *no man would believe in it*.

The report goes on: *This red shift has been analyzed by Dr. Hubble as the speed of distant island universes receeding from the earth, some at 7200 miles a second. Based upon this observation, Dr. Tolman of Pasadena, suggested an expanding universe.* Well, the expanding universe was suggested by Lemaître in 1927. Tolman burst onto the stage in 1929 with several publications [Nussbaumer and Bieri 2009, chapter 13]. In the beginning he remained within de Sitter's frame. In 1930 he threw in the idea of annihilation of matter in order to create a non-static line element in Einstein's formalism. Later in 1930 he became aware of Friedman and Lemaître, and he conceded that a non-static line element could be obtained from an expanding universe without introducing annihilation. Thus, Tolman was an active participant in the cosmological debate, but at the beginning of 1931 he had not yet made up his mind about the appropriate model.

The report continues: *"A theory of an expanding universe," said Dr. Einstein, "at the rate figured from apparent velocities of recession of nebulae would give too short a life to the great universe. It would only be ten thousand million years old, which is altogether too short a time. By that theory it would have started from a small condensation of matter at that time."*

The age problem was indeed a serious challenge. If redshifts were seen as testimonials for the expansion of a finite, spherical universe of radius $R$, and if this expansion was extrapolated backwards to a radius $R=0$, then a lifetime of approximately $10^{10}$ years or shorter was obtained. However, around 1931 the age of the sun and other stars was thought to be much higher than $10^{12}$ years [Nussbaumer and Bieri 2009, chapter 15]; it was based on the erroneous assumption that in stellar interiors the mass of a particle was fully converted to energy and that the total energy available to a star of mass $m$ was therefore $E=mc^2$. Thus the $10^{10}$ years were devastatingly low. This concern was, of course, also on the minds of Eddington and Lemaître. Eddington's solution resided in the long period available to a static Einstein-universe before it slid into expansion. Lemaître's solution was a long period of stagnation after the "Big Bang", when gravitational braking and acceleration driven by the cosmological term, λ, were of nearly equal strength, before λ would push the universe into an accelerated expansion; for details see [Nussbaumer and Bieri 2009, chapter 17]. Einstein knew about Eddington's explication, but Lemaître's "Big Bang" would only be published in the middle of 1931.

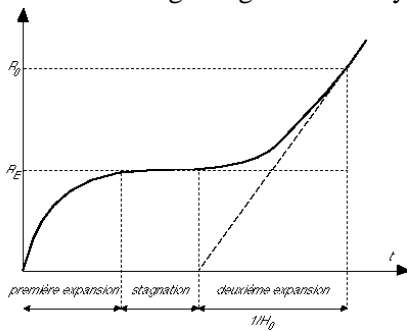

**Lemaître's suggestion of the expansion history.** After the initial decay of the all-containing atom (l'atome primitif) cosmic history was dictated by the relative strength of gravitational attraction and repulsion represented by the cosmological term λ. (From a lecture by Lemaître [Lemaître 1947].

We return to Einstein's further explanations. *He said the red shift might be interpreted as the light quanta getting redder by losing energy as they went long distances. "But no man can get a picture of how this happens," he said.* The "tired light" hypothesis was suggested by Zwicky in 1929 as a possible explanation for the observed redshifts [Zwicky 1929]. Einstein rejects this explanation.

The report goes on: *Dr. Tolman suggested that if we had a periodic solution of contraction and expansion it might be satisfactory.*
　　*Dr. Einstein replied that the equations do not satisfy such a thing, but indicate if such were the case the whole universe might explode, "going swish," he said, laughing.*
　　*"I don't know the answer," said Dr. Einstein.* – At this point the report ends.

This passage shows that in February 1931 Einstein did not realise – or had forgotten – that Friedman gave the solution of a pulsating universe at the end of his article [Friedman 1922]. But two months later Einstein would adopt Friedman's model.

"I don't know the answer" is Einstein's conclusion on February 11, 1931. We are no further than on his arrival on January 3, where he had already accepted, that redshift observations suggested a dynamical universe in contrast to his model of 1917. It is quite clear from both interviews that for him the crucial issue were not distances and redshifts, it was their interpretation, and in the middle of February 1931 his position was: I don't know the answer.

The Einstein Archives contain a draft of a farewell address to the Pasadena community from 1 March 1931, where he thanks them for having been given the opportunity to spend two months with them [Einstein 1931c]. He thanks the "excellent scientists in the fields of physics and astronomy, who showed him their work". *Sie führten mich nicht nur in die Welt der Atome und Krystalle sondern auch an die Oberfläche der Sonne und in die Tiefen des Weltraumes. Ich sah Welten, die sich mit unvorstellbarer Schnelligkeit von uns entfernen, trotzdem ihre Bewohner uns gewiss nicht nahe genug kannten, als dass ein solches Verhalten gerechtfertigt erscheinen könnte. Bei den zahlreichen Ausflügen nach der Sonnenoberfläche erhielt ich (dank St. Johns und Hales vorsichtiger Führung) keinerlei Verbrennungen, bis auf ganz unerhebliche Verbrennungen des Schnabels, die auf die allzu schnelle und unbedenkliche Aeusserungen von Meinungen zurückzuführen waren.* (They not only led me into the world of atoms and crystals but also to the surface of the sun and into the depth of the universe. I saw worlds which withdraw from us at an unimaginable speed, although their inhabitants certainly do not know us sufficiently well that such behaviour could seem to be justified. On the numerous excursions to the surface of the sun I did not suffer any sunburn, thanks to St. John's and Hale's careful guidance, except for some minor burnings of the pecker, which are due to too quick and carelessly expressed opinions.) He then thanks Fleming and Millikan who protected him like two fire-breathing dragons against journalists and others.

For those who claim that meeting Hubble triggered a St. Paul-like conversion in Einstein's cosmology, it must be a sobering disappointment that Hubble was not singled out for special praise, and that his name was not even mentioned in the acknowledgment, in contrast to the solar observers Hale and St. John. – As said before, Einstein's interest in St. John and his solar observations were due to St. John's efforts to find Einstein's gravitational redshift in the solar spectrum. It seems that he spent a lot of time with the solar group.

**8. Einstein's letter to Besso**

The next document which holds a clue to Einstein's conversion is a letter to his friend Besso [Einstein 1931d], dated March 1, 1931. The astronomically relevant passage reads:
*Die Leute vom Mount Wilson-Observatorium sind ausgezeichnet. Sie haben in letzter Zeit gefunden, dass die Spiralnebel räumlich annähernd gleichmässig verteilt sind und einen ihrer Distanz proportionalen mächtigen Dopplereffekt zeigen, der sich übrigens aus der Relativitätstheorie zwanglos folgern lässt (ohne kosmologisches Glied). Der Hacken ist aber, dass die Expansion der Materie auf einen zeitlichen Anfang schliessen lässt, der $10^{10}$, bzw. $10^{11}$ Jahre zurückliegt. Da eine anderweitige Erklärung des Effektes auf grosse Schwierigkeiten stösst, ist die Situation sehr aufregend.* (The people from Mount Wilson-Observatory are excellent. They have lately found that the spiral nebulae are spatially nearly evenly distributed and that they show a Doppler-shift proportionally growing with their distance, which, by the way, follows without problem from the theory of relativity (without the cosmological term). The snag is that the expansion of matter points to a beginning in time which lies $10^{10}$, respectively $10^{11}$ years in the past. As any other explanation meets with great difficulties, the situation is very exciting).

Thus, between the discussion meeting of February 11, reported in the New York Times, and his return to Europe, early March 1931, Einstein's readiness to accept an expanding universe has increased, particularly as he realised that it could be had without the cosmological constant, λ. This suggests that he had had another look at the *Zeitschrift für Physik* (certainly available at Caltec), to study Friedman's article of 1922 in a more receptive frame of mind. There, on the last pages, Friedman presents his periodic model with λ=0. In addition, Lemaître's 1927 paper provided the connection between the increasing radius of curvature and the redshifts in the spectra. Combined with Tolman's hint of a periodic universe, this may have triggered his shift from "I don't know the answer" of February 11 to his much more positive postcard to Besso, where he accepts that the redshifts result naturally from general relativity without having to invoke a cosmological constant. Also the adoption of Friedman's quasi-periodic solution which will come in April 1931, after its rejection at the meeting on February 11, points to a re-evaluation of Friedman at some time between these two dates.

Still, the postcard does not sound like a conversion out of deep conviction, but rather like an acceptance due to a lack of better redshift-explanations, and the "low age" of the resulting world-model still bothers him.

**9. The final step of Einstein's conversion**

In the middle of March 1931 Einstein was back in Berlin. We open his diary [Einstein 1930/1931], which, after an interruption of ten weeks, he resumed on April 8, 1931. An article for Maxwell's 100th anniversary and work on his field equations are on the agenda. Then, on April 13:
*Gestern Vormittag Brahmsquartett mit Frl. Hermann, Frl. Schulz und Cellist Stegmann. Auch Mozart-Divertimento. [...] Nach dem Abendessen Haydn-Trio. Abends im Studierzimmer interessante Idee zum kosmologischen Problem. Heute Korrespondenzen mit Dukas Gesuch an Masaryk zugunsten eines Militärdienst-Verweigerers. Abhandlung zum kosmologischen Problem begonnen.* (Yesterday morning Brahms quartet with Miss Hermann, Miss Schulz and cellist Stegmann. Also Mozart-divertimento. [...] After dinner Haydn-trio. In the evening in my study interesting idea about the cosmological problem. Today correspondence with Dukas petition to Masaryk in favour of a conscientious objector. Began paper on the cosmological problem.)
*16. [...] Nachmittag Akademie. Kosmolog. Arbeit eingereicht. [...].* (Afternoon Academy. Cosmological work submitted.) – This was his publication "Zum kosmologischen Problem der allgemeinen Relativitätstheorie". (About the cosmological problem of general relativity.) [Einstein 1931e].

His report to the Academy fills three pages; let us look at the individual passages. *Unter dem kosmologischen Problem wird die Frage über die Beschaffenheit des Raumes im grossen und über die Art der Verteilung der Materie im grossen verstanden, wobei die Materie der Sterne und Sternsysteme zur Erleichterung der Übersicht durch kontinuierliche Verteilung der Materie ersetzt gedacht wird. Seitdem ich kurz nach Aufstellung der allgemeinen Relativitätstheorie dieses Problem in Angriff nahm, sind nicht nur zahlreiche theoretische Arbeiten über diesen Gegenstand erschienen, sondern es sind durch HUBBELS [sic!] Untersuchungen über den Dopplereffekt und die Verteilung der extra-galaktischen Nebel Tatsachen ans Licht getreten, welche der Theorie neue Wege weisen.* (As cosmological problem we define the question about the constitution of space on the grand scale and about the distribution of matter on the grand scale, where in order to facilitate the general survey we consider the matter of stars and stellar systems as replaced by a continuous distribution of matter. Since, shortly after the establishment of the general theory of relativity, I attacked this problem, there not only appeared numerous theoretical works on this topic, but through HUBBEL'S [sic!] investigations about the Doppler-effect and the distribution of extra-galactic nebulae facts appeared that direct theories into new directions.)

After this introduction Einstein lists his further assumptions of 1917: We live in a homogenous universe, the spatial structure as well as the density remain constant in time. *Damals zeigte ich, dass man beiden Annahmen gerecht werden kann, wenn man das sogenannte kosmologische Glied an die Feldgleichungen der allgemeinen Relativitätstheorie einführt, [...].* (At that time I showed that both assumptions can be satisfied, if the so called cosmological term is introduced into the field equations of general relativity). He then gives equation (1), which we know from his pioneering publication of 1917.

He continues: *Nachdem nun aber durch HUBBELS [sic!] Resultate klar geworden ist, dass die ausser-galaktischen Nebel gleichmässig über den Raum verteilt und in einer Dilatationsbewegung begriffen sind (wenigstens sofern man deren systematische Rotverschiebung als Dopplereffekt zu deuten hat), hat die Annahme (2 [räumliche Struktur und Dichte bleiben constant]) von der statischen Natur des Raumes keine Berechtigung mehr, und es entsteht die Frage, ob die allgemeine Relativitätstheorie von diesen Befunden Rechenschaft zu geben vermag.* (As it has become clear from Hubbel's [sic!] results that the extra-galactic nebulae are spatially evenly distributed and involved in a dilatational motion (at least if their systematic redshifts have to be interpreted as Doppler-shifts) the assumption (2 [spatial structure and density remain constant]) of a static nature of space is no longer justified and the question arises, whether general relativity can account for these features.)

He refers to Hubble's observations: If his redshifts have to be interpreted as Doppler-shifts, then a static universe is no longer tenable and we have to ask, whether general relativity can account for these findings. Einstein seems to have forgotten that general relativity, in the hands of Friedman and Lemaître, had provided models of the expanding universe long before Hubble published about redshifts. Indeed, Lemaître had derived the numerical value of the Hubble constant two years before Hubble [Lemaître 1927, Hubble 1929], and Lemaître's velocity-distance relationship, $v = H \cdot d$, had to wait two years before being observationally confirmed by Hubble. Einstein puts the cart before the horse. But then he continues and remembers that Friedman had done the job, "uninfluenced by observations":
*Es ist von verschiedenen Forschern versucht worden, den neuen Tatsachen durch einen sphärischen Raum gerecht zu werden, dessen Radius P zeitlich veränderlich ist. Als Erster und unbeeinflusst durch Beobachtungstatsachen hat A. Friedman diesen Weg eingeschlagen, auf dessen rechnerische Resultate ich die folgenden Bemerkungen stütze.* (Several investigators have tried to cope with the new facts by using a spherical space whose radius, P, is variable in time. The first who, uninfluenced by observations, tried this way was A. Friedmann, on whose calculations the following remarks will be based). – Einstein contradicts himself: Friedman did not try "to cope with the new facts", but went on his way "uninfluenced by observations". Also Lemaître was most probably originally driven by his interest in finding cosmological solutions to Einstein's field equations (see [Lemaître 1925]), however he knew of Slipher's redshifts, and about the discussion surrounding their interpretation within de Sitter's world model.

Einstein then recalls the form of Friedman's line element (see equation (3)) and the equations for the time variable density and radius of the spherical world (equations (4) and (5)), and he adds: *Aus diesen Gleichungen erhält man meine frühere Lösung, indem man P [Einstein's notation for the radius of curvature] als zeitlich constant voraussetzt. Mit Hilfe dieser Gleichungen lässt sich aber auch zeigen, dass diese Lösung nicht stabil ist, d.h. eine Lösung, welche sich von jener statischen Lösung zu einer gewissen Zeit nur wenig unterscheidet, weicht im Laufe der Zeit immer stärker von jener Lösung ab. Schon aus diesem Grunde bin ich nicht mehr geneigt, meiner damaligen Lösung eine physikalische Bedeutung zuzuschreiben, schon abgesehen von Hubbels [sic!] Beobachtungsresultaten."* (From these equations one obtains my earlier solution, if one assumes *P* [Einstein's notation for the radius of curvature] to be constant in time. With the help of these equations it can be shown that this solution is not stable, i.e. a solution which at a given time only slightly differs from the static solution, increasingly deviates from the static solution. Already on this account I am no longer inclined to ascribe a physical significance to my former solution, quite apart from Hubbel's [sic!] observational results.) And he continues: *Unter diesen Umständen muss man sich die Frage vorlegen, ob man den Tatsachen ohne die Einführung des theoretisch ohnedies unbefriedigenden λ-Gliedes gerecht werden kann* (Under these circumstances one has to ask the question, whether one could not do justice to the facts without introducing the theoretically anyway unsatisfactory λ-term).

This key passage gives us deeper insight into Einstein's motive for abandoning his static world. He refers to Friedman's equations (4) and (5), and says that with their help one can show that his solution of 1917 is not stable, and he adds: "Already on that account I am no longer inclined to ascribe a physical significance to my former solution, quite apart from Hubbel's observational results". Thus, the decisive reason for his change of mind is the instability of his 1917-model. Hubble's observations support his decision, but they are of secondary importance. – This passage argues again for the hypothesis that Eddington's instability-proof triggered Einstein's conversion. In addition, Eddington could confront Einstein with Hubble's paper of 1929, which gave observational support, and was, moreover, endorsed by de Sitter. Hubble's observational material was basically the same as had already been employed by Lemaître. This reminds us that on his arrival in Pasadena Einstein mentioned Lemaître as one of the theoreticians who, concerning the observations of Hubble and Humason, *show a view that fits well into the general theory of relativity*.

Einstein then has to choose a cosmological model. He adopts Friedman's oscillating solution of a positively curved universe, with vanishing cosmological constant, λ=0, and describes the change of the world-radius (the radius of curvature) during one single period. He took his model directly from Friedman [Friedman 1922], who at the end of his paper, when discussing the periodic universe, says: *Setzen wir λ=0 und M= 5·10$^{21}$ Sonnenmassen, so wird die Weltperiode von der Ordnung 10 Milliarden Jahren. Diese Ziffern können aber gewiss nur als Illustration für unsere Rechnungen gelten.* (If we assume λ=0 and M= 5·10$^{21}$ solar masses, then the world-period is of the order of 10 billion years [10$^{10}$ years]. These numbers can certainly only be an illustration of our calculations.) Friedman did not give a source for this number.

In this Friedman-Einstein-universe the radius grows with decreasing expansion velocity from zero to a maximum value and then contracts towards a singularity, which in Friedman's mind was certainly a point where the density was sufficiently high to turn into the next expansion [Friedmann 1923]. Einstein restricts himself to one period.

In order to relate to observations, Einstein defines $D= (R'/R)/c$, where in today's notation $R'/R$ is the Hubble constant $H$, and in an approximate treatment he derives from equation (5) the relation $D^2 \approx \kappa\rho$. From this he obtains an approximate very high mean density of the order of $\rho \approx 10^{-26}$ g/cm$^3$ and an age of the universe of $10^{10}$ years, thus the same age as given by Friedman. About the density he remarks: "*was zu den Abschätzungen der Astronomen einigermassen zu passen scheint*" (which seems to agree tolerably with the estimates of the astronomers). He obviously had not read de Sitter's publication of June 24, 1930, where he discussed Lemaître's solution and where he gave a density of $3.73 \cdot 10^{-29}$ g/cm$^3$ as a compromise between $2 \cdot 10^{-28}$ and $2 \cdot 10^{-30}$ g/cm$^3$ [de Sitter 1930c]. Einstein does not tell us from where he got the observations which enter $D$ nor how he arrived at his age of $10^{10}$ years. O'Raifeartaigh and McCann [2014] suggest that there is an error in Einstein's numerical calculation of the matter density. They also suggest that Einstein's age of the universe is taken from Friedman and repeats an error of Friedman's. Indeed, Einstein's numbers are inconsistent. However, these inconsistencies do not invalidate his argument; had he done the calculations properly with the then accepted $H \approx 500$ (km/s)/Mpc, he would have landed in an even worse age-dilemma of only $\approx 10^9$ years. According to the opinion of the 1930s, it would in any case have lifted the age of the sun above the age of the universe. This problem he declares as *die grösste Schwierigkeit der ganzen Auffassung* (the biggest difficulty of the whole concept). – Belenkiy, when discussing his equations (11) and (12), points to a possible error of the same sort in Friedman's calculation of the age of the universe [Belenkiy 2013], and we have seen that Einstein in his reasoning did heavily lean on Friedman.

Einstein, perhaps with a Haydn-inspired flash of intuition, found a way out of the age dilemma: *Hier kann man der Schwierigkeit durch den Hinweis darauf zu entgehen suchen, dass die Inhomogeneität der Verteilung der Sternmaterie unsere approximative Behandlung illusorisch macht. Ausserdem ist darauf hinzuweisen, dass wohl kaum eine Theorie, welche HUBBEL'S [sic!] gewaltige Verschiebung der Spektrallinien als Dopplereffekte deutet, diese Schwierigkeit in bequemer Weise wird vermeiden können*. (Here one can try to escape from the difficulty by pointing out, that the inhomogeneity in the distribution of stellar matter can render our approximate treatment illusory. Furthermore it has to be pointed out, that probably no theory which interprets Hubbel's [sic!] enormous shifts of the spectral lines as Doppler effect, will be able to avoid this difficulty in a comfortable way.) This "deus ex machina" helped him to overcome the last mental barrier against Lemaître's "abominable" expanding universe. – We have mentioned before, how Eddington and Lemaître dealt with the age problem.

Einstein ended the paper by emphasising that general relativity could account for the observational facts without the cosmological term.

Einstein's note is shockingly scanty of references. We forgive him for naming only "Hubbel" among all the observers. He misspells Hubble's name systematically. However, we don't want to overstate this fact; he simply spelt Hubble's name as a German speaking person might do, if he knew the name from hearing only. But, combined with the fact that Einstein talked about Hubble but gave no reference to any of his papers casts doubt on whether he ever carefully read any of Hubble's publications. For him this "Hubbel" evidently stood for the whole observational community, which had worked on redshifts and distances. His treatment of the theoretical community is much more astonishing. Friedmann (1922) is the only publication he cites. His remark that "Several investigators have tried to cope with the new facts …" is, to put it mildly, a gross understatement. Lemaître had in 1927 given a solution that coped very well with the new facts. And what about his remark that "it can be shown that this solution is not stable"? As mentioned before, Eddington's proof that the static Einstein-universe was unstable was probably the initiating kick on his hesitating path towards the expanding universe; Eddington would certainly have deserved a citation. However, we should not be too harsh on Einstein. He did not claim a new discovery, indeed, his paper contains no new physics, all the cosmological twists had already been discussed by others. The main purpose of his public conversion was to acknowledge that his former solution did not represent a stable state and – even more important – that he had gotten rid of the cosmological constant.

Einstein's paper contains a psychologically interesting passage: "The first who, uninfluenced by observations, tried this way was A. Friedmann, on whose mathematical results the following remarks will be based". It was a belated homage to Friedman, who had theoretically discovered what escaped Einstein in his paper of 1917, and again in 1922 and 1923 when he brushed Friedman's publication aside, and even in 1927 when Lemaître told him about the relevant observations. But it was too late for personal reparation, Friedman had died in 1925.

On June 27, 1931, Einstein sent a letter to Tolman, to which he joined his paper on the cosmological problem [Einstein 1931f]. About his publication he said that he only wanted to point out that the λ-term became unnecessary, if solutions with a time-variable radius of the universe were allowed. *Dies* *ist ja wirklich unvergleichlich befriedigender.* (That is indeed incomparably more satisfactory). Einstein ends his letter with a postscript: *Wie Sie sehen, bin ich wieder davon abgekommen, Hubbels [sic!] Linienverschiebungen in irgenwie abenteuerlicher Weise zu deuten.* (As you can see, I abandoned my attempts to interpret Hubbel's [sic !] lineshifts in some adventurous fashion.)

Tolman answered on September 14. He liked Einstein's quasi-periodic universe, but did not agree with his dismissal of the cosmological constant. After all, λ might be a new constant of nature, and *since the introduction of the λ-term provides the most general possible expression of the second order which would have the right properties for the energy-momentum tensor, a definite assignment of the value λ=0, in the absence of an experimental determination of its magnitude, seems arbitrary and not necessarily correct* [Tolman 1931]. – In their correspondence Einstein wrote in German and Tolman in English; it may well be that their spoken conversation proceeded in the same way.

Tolman's attitude about the cosmological constant was shared by others, in particular by Eddington and Lemaître. Both were convinced that λ represented a new force in nature, a force of the same importance as gravitation.

## 10. Towards the Einstein-de Sitter universe

The political climate in Germany deteriorated further; at the beginning of December Einstein was on board the "Portland", sailing for America. On December 6, 1931 he confides to his diary: *[…] Heute entschloss ich mich, meine Berliner Stellung im Wesentlichen aufzugeben. Also Zugvogel für den Lebensrest!* (Today I decided to essentially give up my position in Berlin. A vagrant bird for the rest of my life!) [Einstein 1931g].

On December 30, 1931 the ship arrived in Los Angeles; in Einstein's diary we read:
*Ankunft in Los Angeles spät abends am 30. XII. Viele Kriegsschiffe im Hafen, Lichtermeer. Am nächsten Morgen von Tolmann beim Schiff abgeholt. Vorläufig einquartiert bei Fleming wo wir 5 Tage blieben.* (Many war-ships in the port, immense multitude of lights. Next morning Tolman fetches me on board. Temporarily accommodated at Flemings home, where we stayed for 5 days.)

*8. I. Abfassung einer Notiz mit De Sitter über Beziehung zwischen Hubbel-Effekt [sic!] und Materie-Dichte der Welt bei Vernachlässigung der Krümmung und des λ-Gliedes.* (Composition of a note with De Sitter about the relationship between Hubbel-effect [sic!] and matter-density, neglecting curvature and λ-term.)

*25. Grosses Dinner mit Rede von mir Aermsten. Tolman übersetzt sie. Rechnen mit Tolman über kosmologisches Problem.* (Large dinner with a speech by poor me. Tolman translates. Calculation with Tolman on the cosmological problem.) – This shows again the friendly relationship between Einstein and Tolman. And also in this diary I looked in vain for Hubble's name, although he talks about astronomical seminars, talks by various academics, and social engagements of all kind.

Thus, on January 8, 1932 Einstein and de Sitter teamed up to invent what would become the Einstein-de Sitter universe. Their message covered two pages and had two focal points: the elimination of the cosmological constant, and a shift to zero-curvature. *There is no direct observational evidence for the curvature, the only directly observed data being the mean density and the expansion, which latter proves that the actual universe corresponds to the non-statical case. It is therefore clear that from the direct data of observation we can derive neither the sign nor the value of the curvature, and the* *question arises whether it is possible to represent the observed facts without introducing a curvature at all* [Einstein and de Sitter 1932]. They referred to a paper by Heckmann, who had demonstrated that Lemaître's solutions of Einstein's field equations were also valid for an Euclidean universe [Heckmann 1931]; actually, the case of an Euclidean universe had already been treated by Lemaître in 1925.

They quite unceremoniously eliminated the cosmological constant $\lambda$:
*Historically the term containing the "cosmological constant" $\lambda$ was introduced into the field equations in order to enable us to account theoretically for the existence of a finite mean density in a static universe. It now appears that in the dynamical case this end can be reached without the introduction of $\lambda$.*

Having stated their two principle messages, they formulated their new model. Opting for zero curvature, their universe can be described in Euclidean geometry. Friedman's line element (3) takes the form

$$ds^2 = -R^2\left(dx_1^2 + dx_2^2 + dx_3^2\right) + c^2 dt_4^2 \quad (7)$$

where *R* depends on time only. From Friedman's equations (4) and (5) they needed only the one that provides the relation between the parameter of expansion, *R*, the coefficient of expansion, *H*, and the mean density $\rho$. They found

$$\frac{1}{R^2}\left(\frac{dR}{cdt}\right)^2 = \frac{1}{3}\kappa\rho, \quad (8)$$

where

$$\frac{1}{R}\left(\frac{dR}{cdt}\right) = H. \quad (9)$$

In principle, the relationship $H^2 = \frac{1}{3}\kappa\rho$ is easy to verify by observation. In those days it was generally accepted that $H \approx 500$ (km/s)/Mpc. This resulted in a mean density of $\rho = 4 \cdot 10^{-28}$ g/cm$^3$, which happened to coincide with the observational upper limit determined by de Sitter on earlier occasions, but they hasten to add that observational uncertainties were high, implying accordingly high uncertainties in *H* and $\rho$.

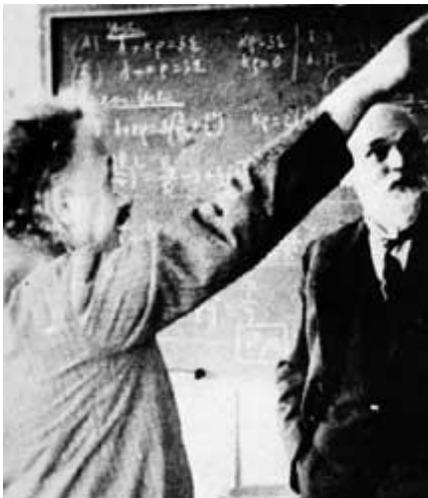

**Einstein and de Sitter.** Photographed January 8, 1932 at Pasadena, California Institute of Technology. PHOTO: LEIDEN OBSERVATORY ARCHIVES.

The Einstein-de Sitter universe became the favourite cosmological model. Whereas the authors made short work of the cosmological constant, λ, they left the door open to non-zero curvature: *The curvature is, however, essentially determinable, and an increase in the precision of the data derived from observations will enable us in the future to fix its sign and to determine its value.* – Observationally their model would hardly be distinguishable from Lemaître's finite, spherical model in the case of λ=0, and $r<<R$ for any observable distance $r$.

Thus, the two scientists who had been the first to propose cosmological models, but later saw them drown, had joined forces to propose a new version which was quickly accepted by a large fraction of the scientific community. Eddington and Lemaître opposed the banishment of λ. Eddington's strong defence of λ is unequivocally formulated in his address to the 1932 meeting of the IAU: *I would as soon think of reverting to Newtonian theory as of dropping the cosmological constant* [Eddington 1933].

Einstein and de Sitter had to admit a discrepancy between the observational and their theoretical ratio $H^2/\rho$. For a long time this was blamed on observational uncertainties. Towards the end of the twentieth century observations of the time-dependent expansion rate of the universe, determined from supernovae, made it clear that the Einstein-de Sitter cosmology had to be abandoned. In 2011 these observations were acknowledged with the Nobel Prize. They showed plausibly that the expansion velocity of the universe did not obey the behaviour expected from the Einstein-de Sitter model, but rather supported the Lemaître-hypothesis, where at some time in the past the cosmological term began to dominate over gravitational braking of the expansion. The cosmological constant, λ, frowned upon by Einstein, yet beloved by Lemaître and Eddington, was allowed back into cosmology. – However, by that time de Sitter and Einstein, as well as Eddington and Lemaître had been dead for many years.

## 11. Conclusions

Einstein left hardly any documentation describing his way from the static to the dynamic universe. We have tried to collect all the written testimonials that could shed light on this episode. When in 1917 Einstein founded modern cosmology, his aim was to describe our universe in terms of general relativity; his assumption of a static universe was based on common sense of that time. In order to obtain that static world he complemented his fundamental equations with the cosmological term, λ, which acted as a counter force to gravitation. De Sitter's competing model, published shortly after Einstein's initial work, was empty of matter, but had the right properties to apparently explain the redshifts of extragalactic nebulae, discovered by Slipher in 1912. De Sitter's cosmology provoked a lively discussion among theoreticians about the appropriate interpretation. It is all the more astonishing how high-handedly Einstein qualified Friedman's work as irrelevant, when he showed that general relativity allowed the existence of a dynamical universe. And we are even more astonished about the vehemence with which Einstein disqualified Lemaître's discovery, whose theoretical arguments were supported by observational evidence. – The immutability of the universe was obviously a very deep-rooted conviction of Einstein.

The available documentation strongly suggests that Einstein's reluctant conversion began in Cambridge, when in June 1930 Eddington confronted him with the fact that the static Einstein-model was unstable. In addition, Eddington certainly informed Einstein about de Sitter's switch of allegiance to Lemaître's expanding universe, and that the dynamical universe had received strong observational backing from Hubble's publication of 1929, which de Sitter had verified in 1930.

When in December 1930 Einstein finally followed a long standing invitation of Millikan to visit Caltech, he already knew about Hubble's observational discoveries and Lemaître's hypothesis of an expanding universe, and he knew that Caltech's Tolman was involved in the cosmological debate. All this is clear from his January 2, 1931 New York Times interview. There is no evidence that Hubble and Einstein indulged in any profound discussions, which would have influenced the latter's cosmological concepts.

The available information strongly contradicts a popular cliché, which claims that Einstein was converted to the expanding universe by Hubble, when he showed him his observations in January 1931. On the other hand, in Pasadena Einstein also met Tolman, who at the time was very active in theoretical cosmology, and they certainly discussed these matters at some length. From the Los Angeles Times of February 5, 1931 we could suspect that for a short while Einstein even hoped that the unified field theory might show a way out of the cosmological enigma. However, up to February 11 Einstein had not yet made up his mind, how to switch from his former firm belief in a static world to a dynamic model, as demanded by theory and observation. He had in front of him Lemaître's expanding, λ-driven, spherical universe, which explained Hubble's observational results. He also had Friedman's theoretical models, offering an extended choice of evolutionary paths.

The letter to Besso shows that during the second half of February 1931 Einstein must have had another look at some cosmological key-publications, in particular at Friedman's paper of 1922. On March 1, he seems inclined to accept a dynamic universe. However, there remained a last obstacle. The age of the universe, calculated from the redshifts of the nebulae and interpreted in the framework of the relativistic expansion theory, was much shorter than the lifetime of stars, as believed in those days. After his return to Berlin Einstein found a way out of that dilemma by invoking inhomogeneities in the distribution of matter; they might account for the "low age" resulting from the model based on a homogeneous density distribution. This loophole must have given him the courage to opt for Friedman's periodic universe with λ=0, thus freeing himself at the same time from the cosmological term, to which he had never warmed up.

Einstein's publication of 1931 was scientifically irrelevant, it contained nothing new. However, it is a historical landmark, as it announced Einstein's conversion to a dynamic, expanding universe without the action of a cosmological constant.

In January 1932 Einstein was back in Pasadena. There he met de Sitter. Together they published the Einstein-de Sitter universe, which became the standard model up to the middle of the 1990s. It is a spatially flat, ever expanding universe with λ=0 and an expansion velocity asymptotically approaching zero in the infinite future. Whereas the authors admit the possibility that better observations might bring curvature back into cosmology, they kicked the cosmological constant completely out of their equations; on this point history has proven them wrong.

*Note added in proof.* Document 2-112 of the Albert Einstein Archives is a draft "Zum kosmologischen Problem", handwritten by Einstein on American paper and assigned by AEA to 1931. Its content dates it to January or February. Einstein proposes a line element $ds^2 = -e^{\alpha t}\left(dx_1^2 + dx_2^2 + dx_3^2\right) + c^2 dt^2$. Such a time-dependent factor $e^{\alpha t}$ had already been introduced by Tolman in his 1930 papers, where α served as a measure for the annihilation of matter. Einstein now assigns to α the role of a creation source. He changes his former assumption of a static universe to an expanding one of constant density ρ. After coordinate transformation his fundamental equations provide him with two relations between α, ρ, and λ: α=α(ρ,λ), and he obtains $α^2=κc^2ρ/3$, κ=Einstein's constant of gravity. To maintain a constant ρ, the outflowing particles need to be continually replaced by creating new ones. He thought the law of conservation respected, as with λ "space itself is not energetically empty". This gave him an expanding, spatially Euclidean, steady-state-universe. However, when deriving α=α(ρ,λ) he had introduced a numerical error. When he subsequently corrected the error his two equations only yielded a solution for ρ=0: $α^2=4λc^2/3$, taking him basically back to de Sitter's empty universe.

The draft was probably intended as a short paper and discussed with Tolman. We now recall Einstein's diary of January 3: "… but Tolman is right". We may speculate that Tolman found Einstein's numerical error and thus aborted his attempt to obtain a time-dependent line element, which would provide the observed redshifts. There are no indications that Einstein pursued the idea of a steady state universe any further. But this scrap of paper shows that at the beginning of 1931 he was quite prepared to keep λ in his repertory and assign it a new role.

**Acknowledgements.** This article draws heavily on "Discovering the Expanding Universe" by Lydia Bieri and myself [Nussbaumer and Bieri 2009]. I thank Lydia Bieri, David Topper, and Michael Way for comments and suggestions concerning this paper. I thank Norbert Straumann for having drawn my attention to Einstein's considering the λ-term already in 1916. Many thanks to Barbara Wolff of the Albert Einstein Archives of Jerusalem for help in the search for additional documents, and to Cormac O' Raifeartaigh for provoking me to write this article.